\documentclass[cmp]{mysvjour}
\def\C{\mathbb C}

\def\P{\mathbb P}
\def\R{\mathbb R}

\def\Y{\mathbb Y}
\def\Z{\mathbb Z}

\def\ffi{{\varphi}}

\def\wt{\widetilde}

\def\cA{{\rm A}}
\def\cB{{\rm B}}
\def\ccB{{\mathcal B}}

\def\cD{{\mathbb D}}

\def\cJ{{\mathcal J}}

\def\cS{{\mathcal S}}
\def\cT{{\rm T}}
\def\cR{{\rm R}}

\def\dist{{\rm{dist}}}

\def\u0{{\mathbf 0}}

\def\uu{{\mathbf u}}
\def\uv{{\mathbf v}}

\def\ux{{\mathbf x}}
\def\uy{{\mathbf y}}

\def\uH{{\mathbf H}}
\def\uG{{\mathbf G}}
\def\uH{{\mathbf H}}
\def\uS{{\mathbf S}}

\def\rx{{{\rm x}}}
\def\ry{{{\rm y}}}
\def\ru{{{\rm u}}}

\def\wt{\widetilde}
\def\eps{{\epsilon}}
\def\om{{\omega}}
\def\Lam{{\Lambda}}
\def\Bphi{{\mbox{\boldmath${\phi}$}}}

\def\BPsi{{\mbox{\boldmath${\Psi}$}}}

\def\BLam{{\mbox{\boldmath${\Lam}$}}}

\def\lam{{\lambda}}
\def\half{\frac{1}{2}}

\def\pt{\partial}
\def\pn{\par\noindent}
\def\pmn{\par\medskip\noindent}
\def\psn{\par\smallskip\noindent}
\def\z2{{\Z^2}}
\def\zp2{{\Z^2_{\geq}}}

\def\myset#1{{\left\{\,#1\,\right\}}}

\def\pr#1{{  \P\left\{ \, #1 \, \right\}  }}

\def\card{{\,{\rm card }\,}}
\def\dist{{\,{\rm dist}}}

\def\myproof#1{{\pmn{\it Proof of #1. }}}

\def\truc#1#2#3{\smash{\mathop{\,\, #1 \,\, }\limits^{#2}_{#3}}}

\def\elli{{\ell}}

\def\DSzero{{\bf (DS.{\mbox{\boldmath${0, I, N}$}})}}

\def\DSk{{\bf (DS.{\mbox{\boldmath${k, I, N}$}})}}
\def\dsk#1{{\bf (DS.{\mbox{\boldmath${k, I, {#1}}$}})}}

\def\DSkone{{\bf (DS.{\mbox{\boldmath${k+1, I, N}$}})}}

\def\Szero{{\bf (S.{\mbox{\boldmath${0, I, N}$}})}}

\def\ISk{{\bf (IS.{\mbox{\boldmath${k.N}$}})}}

\def\ISkone{ {\bf (IS.{\mbox{\boldmath${k+1.N}$}})} }

\usepackage{graphics}

\usepackage{latexsym}
\usepackage{amsfonts, amssymb}
\usepackage{amsmath}
\usepackage{graphicx}
\usepackage{array}
\usepackage{rotating}

\spnewtheorem{Thm}{Theorem}[section]{\bf}{\it}
\spnewtheorem{Lem}{Lemma}[section]{\bf}{\it}
\spnewtheorem{Def}{Definition}[section]{\bf}{\it}
\spnewtheorem{Cor}{Corollary}[section]{\bf}{\it}

\numberwithin{equation}{section}

\begin{document}

\title{Multi-Particle Anderson Localisation: \\ Induction on the Number of Particles}

\author{Victor Chulaevsky \inst{1} \and Yuri Suhov\inst{2}
}                     
\institute{D\'{e}partement de Math\'{e}matiques et d'Informatique,
Universit\'{e} de Reims, Moulin de la Housse, B.P. 1039, 51687 Reims Cedex 2, France;
\email victor.tchoulaevski@univ-reims.fr 
\and  Department of Pure Mathematics and Mathematical
Statistics, University of Cambridge, Wilberforce Road, Cambridge CB3 0WB, UK}
\date{Received: date / Accepted: date}
\date{\today}
%



\maketitle

\begin{abstract}
This paper is a follow-up of our recent papers \cite{CS08} and \cite{CS09}
covering the two-particle Anderson model. Here we
establish the phenomenon of  Anderson localisation for a quantum
$N$-particle system on a lattice $\Z^d$ with short-range interaction
and in presence of an IID external potential with sufficiently
regular marginal cumulative distribution function (CDF).
Our main method is an adaptation of the multi-scale analysis
(MSA; cf. \cite{FS}, \cite{FMSS}, \cite{DK}) to multi-particle systems,
in combination with an induction on the number of particles, as was proposed in
our earlier manuscript \cite{CS07}. Similar
results have been recently obtained in an independent work by
Aizenman and Warzel \cite{AW08}: they proposed an extension of the
Fractional-Moment Method (FMM) developed earlier for single-particle
models in \cite{AM93} and \cite{ASFH01} (see also references therein)
which is also combined with an induction on the number of particles.

An important role in our proof is played by a variant of Stollmann's eigenvalue concentration bound (cf. \cite{St00}). This result, as was proved earlier in \cite{C08}, admits a straightforward
extension covering the case of multi-particle systems with correlated external random potentials:
a subject of our future work. We also stress that the scheme of our proof
is \textit{not} specific to lattice systems, since our main method, the MSA, admits a continuous version. A proof of multi-particle Anderson localization  in continuous interacting systems with various types of external random potentials will be published in a separate papers.
\end{abstract}

\section{Introduction and the main result }
\label{intro}

The status of the multi-particle Anderson localisation problem has been
described in \cite{AW08}, Section 1.1; the reader is advised to consult
this reference.

The configuration space of the $N$-particle lattice system is the Cartesian
product $\Z^d\times\cdots\times\Z^d$ of $N$ copies of a cubic lattice $\Z^d$,
which we denote for brevity by $\Z^{Nd}$. The
Hilbert space of the $N$-particle lattice system is
$\ell_2(\Z^{Nd})$. The Hamiltonian $\uH \left(=\uH^{(N)}_{U,V,g} (\om )\right)$
is a lattice Schr\"{o}dinger operator acting on functions
$\Bphi\in\ell_2(\Z^{Nd})$ by
$$
\begin{array}{ll}\uH^{(N)}\Bphi (\ux)&=H^0\Bphi (\ux)+ (U(\ux)+gW(\ux;\om))\Bphi(\ux )\\
\;&\;\\
\;&=\sum\limits_{\uy\in\Z^{Nd}:\atop{\|\uy - \ux\|=1}} \,\Bphi(\uy) + \left[U(\ux)+gW(\ux;\om)\right] \Bphi(\ux),\\
\hbox{where}&W(\ux;\om) =\sum_{j=1}^N V(x_j;\om),\\
\;&\quad\ux=(x_1,\ldots ,x_N),\;\uy=(y_1,\ldots ,y_N) \in\Z^{Nd}.
\end{array}
\eqno(1.1)
$$
Here and below we use boldface letters such as $\ux$, $\uy$, $\uH$,
etc., referring to a multi-particle system, where the particle number
enters as an index or specified verbally. For example, small-case
boldface letters $\ux$, $\uy$, etc., will stand for designate
points in $\Z^{Nd}$, called
$N$-particle configurations. Letters $x$, $y$ will be systematically
used for points in $\Z^d$ or $\R^d$, referred to as single-particle
positions (or briefly, positions).

Our proof of $N$-particle Anderson
localisation is organised as
an induction in $N$, as has been explained in earlier presentations
(see, e.g., \cite{CS07}). Thus, we will have to deal with
systems with smaller number of particles, $1\le n<N$.
The respective objects, viz., points in $\Z^{nd}$, $n<N$,
are still denoted by boldface letters:
$\ux \in \Z^{nd}$, $\uy \in \Z^{nd}$, etc.

Next,  $x_j=\big(\rx_j^{(1)},\ldots,\rx_j^{(d)}\big)$ and
$y_j=\big(\ry_j^{(1)},\ldots, \ry_j^{(d)}\big)$ stand for the
positions of individual particles in $\Z^d$,
$j=1,\ldots ,N$, and $\|\cdot\|$ denotes the sup-norm: for
$\uv = (v_1,\ldots ,v_N)\in\R^d\times\cdots\times\R^d:=\R^{Nd}$,
$$
\| \uv\| =\max_{j=1,2}\; \|v_j\|,
\eqno (1.2.1)
$$
where, for $v=({ \rm v}^{(1)}, \ldots, { \rm v}^{(d)})\in\R^d$,
$$\|v\|= \max_{i=1, \dots, d}\left|{\rm v}^{(i)}\right|.\eqno (1.2.2)$$

We will consider the distance on $\R^{Nd}$, $\Z^{Nd}$ and $\R^d$, $\Z^d$
generated by the norm $\|\;\cdot\;\|$.

Throughout this paper, the random external potential
$V(x;\omega )$, $x\in\Z^d$, is assumed to be
real IID, with a common CDF ${\rm F}_V$ on $\R$.
The condition on ${\rm F}_V$ guaranteeing the validity of
our results is as follows:
$$
\sup_{\eps \in (0,1)}\left[\frac{1}{\eps^A}
\sup_{a\in \R}\;\big({\rm F}_V(a+\eps)-{\rm F}_V(a)\big)\right]
<+\infty,\eqno(1.3)
$$
for some $A>0$. In other words, the marginal distribution of the random potential is
H\"{o}lder-continuous\footnote{One can easily show that the main result of this paper remains valid for
log-H\"{o}lder continuous CDF $F_v$, satisfying $|F_V(a+\eps) - F_V(a)| \le C\, \ln^{-A}|\eps|^{-1}$
with $A>0$ large enough.}. Clearly, this does not require the absolute continuity of $F_V$.

Parameter $g\in\R$ is traditionally called the coupling, or
amplitude, constant.

The interaction energy function $U$ is assumed to be of the form
$$U(\ux )=\sum_{1\leq j_1<j_2\leq N}\Phi (x_{j_1},x_{j_2}),
\;\;\ux =(x_1,\ldots ,x_N)\in\Z^{Nd},\eqno (1.4)$$
where function $\Phi:\;\Z^d\times\Z^d\to\R$ (the two-body
interaction potential) satisfies the following properties.
\psn
(i) {\sl $\Phi$ is a bounded symmetric function:}
$$\sup\;\big[|\Phi (x,x')|:\;x,x'\in\Z^d\big]<+\infty ,\;\;
\Phi (x,x')=\Phi (x',x),\;x,x'\in\Z^d.
\eqno (1.5.1)
$$
(ii) {\sl $\Phi$ has a finite range:}
$$\Phi (x,x')=0,\;\hbox{ if }\;\|x-x'\|>r_0,\eqno(1.5.2)$$
where $r_0\in [0,+\infty)$ is a given value.

It is then obvious that function $U:\;\Z^{Nd}\to\R$ is symmetric
under any permutation of positions $x_j$: $U({\mathbf x})=
U(\cS_\sigma{\mathbf x})$. Here $\sigma$ is an arbitrary
element of the symmetric group ${\mathfrak S}_N$, and,
given $\ux=(x_1, \ldots,  x_N) \in\Z^{Nd}$,
$$
\cS_\sigma\ux= (x_{\sigma(1)}, \ldots,  x_{\sigma(N)}).
$$
The same is true for function $W$ (see Eqn (1.1)).

We consider binary interaction potentials in order not to make our notations excessively
cumbersome. The reader will see that, actually, more general bounded short-range many-body interactions
can be treated in the same way. The symmetry does not play an  important role, but is convenient
technically (and natural from the physical point of view).

Throughout the paper, $\P$ stands for the joint probability
distribution of RVs
$\{V(x;\om), x\in\Z^d\}$. The main assertion of this paper is

\begin{Thm} \label{MThm} Consider the random Hamiltonian
$\uH^{(N)}(\om)$ given by {\rm{(1.1)}}.
Suppose that $U$ satisfies conditions {\rm{(1.4)}} and {\rm{(1.5)}},
and the random potential
$\{V(x;\om)$, $x\in\Z^d\}$ is {\rm{IID}} obeying {\rm{(1.3)}}.
Then there exists
$g^*\in(0,+\infty)$ such that for any $g$ with $|g|\geq g^*$,
the spectrum of operator $\uH^{(N)}(\om )$
is $\P$-a.s. pure point. Furthermore, there exists a nonrandom
constant $m_+ = m_+(g) >0$  such that all eigenfunctions
$\BPsi_j(\ux;\om)$ of $\uH^{(N)}(\om )$ admit an exponential bound:
$$
|\BPsi_j(\ux;\om)|\leq C_j(\om) \, e^{-m_+\|\ux\|}.\eqno(1.6)$$
\end{Thm}

The assertion of Theorem \ref{MThm} can also be stated in the form
where $\forall$ given $m_*>0$, $\exists$ $g_*=g_*(m_*)\in (0,+\infty )$
such that $\forall$ $g$ with $|g|\geq g_*$, the eigenfunctions
$\BPsi_j(\ux;\om)$ of $\uH^{(N)}(\om )$ admit exponential bound (1.6).
\psn
\textbf{Remarks. 1. } The threshold value $g^*$ in Theorem \ref{MThm} depends on $N$: $g^* = g^*(N)$.
(It also depends on $\rm F_V$ and $\Phi$.) The important question is how $g^*$ grows with $N$.
We plan to address this problem in a separate paper.
\psn
{\textbf{2.} It suffices to prove Theorem \ref{MThm} for any bounded interval $I\subset \R$ of length $\ge \delta_0$ with a given, suitably chosen $\delta_0>0$. This is convenient (albeit not crucial) in some arguments used below.
\pmn

The conditions of Theorem \ref{MThm} are assumed throughout the paper.
As was said earlier, the proof of Theorem \ref{MThm}  uses mainly MSA,
in its $N$-particle version. The MSA scheme for $N$ particles does not differ in principle from
that for two particles; for that reason, we will often refer to paper \cite{CS09}.

Most of the time we work with
finite-volume approximation operators $H^{(N)}_{\BLam^{(N)}_{L}(\uu)}\left(=
H^{(N)}_{\BLam^{(N)}_{L}(\uu)} (\om )\right)$ given by
$$
\uH^{(N)}_{\BLam^{(N)}_{L}(\uu)} =
\uH^{(N)}\upharpoonright_{\BLam^{(N)}_L(\uu)}
+ \text{ Dirichlet boundary conditions on }\; \partial\BLam^{(N)}_L(\uu )
\eqno(1.7)
$$
and acting on vectors $\Bphi\in\C^{\BLam^{(N)}_{L}(\uu)}$ by
$$
\begin{array}{r}\uH^{(N)}_{\BLam^{(N)}_L(\uu)}\Bphi (\ux)=
\sum\limits_{\uy\in\BLam^{(N)}_L(\uu ):\atop{\|\uy - \ux\|=1}}
\,\Bphi(\uy)
+  \left[U(\ux)+gW(\ux;\om)\right] \Bphi(\ux),
\end{array}
\eqno(1.8)
$$
with the external $N$-particle random potential $W(\ux;\om)$ as in (1.1). Here and below,
$\BLam^{(N)}_L(\uu)$ stands for an `$N$-particle lattice box' (a box,
for short) of size $L$ around
$\uu=(u_1, \ldots, u_N)$, where $u_j=(\ru^{(1)}_j, \ldots,
\ru^{(d)}_j)\in\Z^d$:
$$\BLam^{(N)}_L(\uu) = \truc{\times }{N}{j=1}\Lam_L(u_j)
\eqno (1.9.1)$$
where $\Lam_L(u_j)$ is a `single-particle box' around
$u_j=\left({\ru}_j^{1)},\ldots ,{\ru}_j^{d)}\right)\in\Z^d$:
$$\Lam_L(u_j)=\left(\truc{\times }{d}{i=1}
\left[\ru_j^{(i)}-L/2, \ru_j^{(i)}+L/2 \right]\right)\cap \Z^{d}.
\eqno (1.9.2)$$
For a box $\BLam^{(N)}_L(\uu)$ as in (1.9.1), we will also use the
notation:
$$\Pi_j \BLam^{(N)}_{L}(\uu)=\Lam_L(u_j)$$
and
$$\Pi \BLam^{(N)}_L(\uu) = \cup_{j=1}^N \Pi_j \BLam^{(N)}_L(\uu);\eqno (1.9.3)
$$
set $\Pi\BLam^{(N)}_L(\uu )\subset\Z^d$ describes the single-particle `base' of
$\BLam^{(N)}_L(\uu )$.

Next, $\pt \BLam^{(N)}_L(\uu)$ in (1.7) stands for the interior boundary
(or briefly, the boundary) of box $\BLam^{(N)}_L(\uu)$:
$\pt \BLam^{(N)}_L(\uu)$
is formed by points $\uy\in\BLam^{(N)}_L(\uu)$ such that $\exists$ a site
$\uv\in\big(\Z^{Nd}\big)\setminus\BLam^{(N)}_L(\uu)$ with $\|\uy -\uv\|=1$.
These definitions remain valid if we replace $N$ with $n=1, \ldots, N-1$.

As follows from (1.7), (1.8),
$H^{(N)}_{\BLam^{(N)}_L(\uu)}$ is a Hermitian operator in the Hilbert
space $\ell_2(\BLam^{(N)}_L(\uu))$
In fact, the approximation (1.7) can be used for any finite
subset $\BLam^{(N)}\subset\Z^{Nd}$ of cardinality $|\BLam^{(N)} |$ and
with boundary $\pt\BLam^{(N)}$,
producing Hermitian operator $\uH^{(N)}_{\BLam^{(N)}}$ in $\ell_2(\BLam^{(N)})$.

Hamiltonian $\uH^{(N)}$ and its approximants $\uH^{(N)}_{\BLam^{(N)}}$
admit the permutation symmetry.
Namely, let $\uS_\sigma$ be the unitary operator in
$\ell_2(\Z^{Nd})$ induced by map $\cS_\sigma$:
$$
\uS_\sigma \Bphi (\ux) = \Bphi(\cS_\sigma\ux ).
\eqno(1.10)
$$
Then $\uS_\sigma^{-1}\uH^{(N)}\uS_\sigma =\uH^{(N)}$ and
$\uS_\sigma^{-1}\uH^{(N)}_{\BLam^{(N)}} \uS_\sigma =\uH^{(N)}_{\sigma\BLam^{(N)}}$.
This implies, in particular, that for
any finite $\BLam^{(N)} \subset \Z^{Nd}$, the eigenvalues
of operators $\uH^{(N)}_{\BLam^{(N)}}$ and $\uH^{(N)}_{\cS_\sigma\BLam^{(N)}}$
are identical.

Like its two-particle counterpart (see \cite{CS08}, \cite{CS09}),
the $N$-particle
MSA scheme involves a number of technical parameters borrowed
from the single-particle MSA; see \cite{DK}. Following \cite{DK} and
\cite{CS08}, \cite{CS09}, given a number
$\alpha\in(1,2)$ and
starting with $L_0\geq 2$ and $m_0>0$, we define an increasing
positive sequence $L_k$:
$$L_k = L_0^{\alpha^k},\;\; k\geq 1,\eqno(1.11)$$
and a decreasing positive sequence $m_k$ (depending on a positive
number $\gamma$):
$$m_k = m_0 \,\prod_{j=1}^k \left( 1 - \gamma L_k^{-1/2} \right)
,\;\; k\geq 1.\eqno (1.12)
$$
In fact, it suffices to set $\alpha=3/2$, albeit we will use the symbolic form of parameter $\alpha$ instead of its value: this makes our notations less cumbersome. Besides, it will make our notation agreed with that of \cite{DK}.

We will also make use of parameters
$$
p=p(N, g)>d  \text{ and } q=q(N,p(N,g)) > p,
\eqno(1.13)
$$
varying with the number of particles $N$. The roles
of parameters $p$ and $q$ (and the choice of their values) have been
discussed in \cite{CS09}: they appear systematically in
the exponents of power-law bounds for probabilities of "unwanted",
or "unlikely" events defined in terms of finite-volume Hamiltonians
$\uH^{(N)}_\BLam$. These bounds also depend on $d$, $\alpha$ and
$\gamma$ (which could be added to the list of arguments for $p$ and $q$)
 and are specified, for a given value of $N$, recursively,
depending on the values $\{p(n)$ and $q(n, p(n))$ for $n$-particle systems,
where $ n=1, \ldots, N-1\}$.
In the course of presentation, it will be made  clear (and used in various places) that, for any $N\geq 1$,
$$
p(n,g), q(n,g) \to +\infty\;\hbox{ as $|g|\to\infty$}, \; n=1, \ldots, N.
\eqno (1.14)
$$

Note that sequence $m_k$ in (1.12) is indeed positive, and the
limit $\lim\limits_{k\to\infty} m_k  \geq m_0/2$ when
$L_0$ is sufficiently large. We will also assume that $L_0>r_0$.
(A similar observation was, in fact, made in the Appendix in \cite{DK}.)

The single-particle MSA scheme was used in \cite{DK} to check,
for IID potentials, decay properties of the Green's functions
(GFs) for single-particle Hamiltonians with IID external potentials. As was said before, for a two-particle model, the MSA scheme was established
in \cite{CS08}, \cite{CS09}. In this paper we adopt a similar
strategy for the
$N$-particle model. Here, the GFs in a box $\BLam^{(N)}_L(\uu )$ are
defined by:
$$
G^{(N)}_{\BLam^{(N)}_L(\uu)}(E; \ux, \uy) =\left\langle \left(
\uH^{(N)}_{\BLam^{(N)}_L(\uu)} - E\right)^{-1}
\delta_{\ux}, \delta_{\uy}
\right\rangle, \; \ux,\uy\in\BLam^{(N)}_L(\uu),\eqno(1.15)$$
where $\delta_{\ux}(\uv )$ is the lattice delta-function and
$\langle \cdot, \cdot \rangle$ stands for the scalar product
in $\ell_2(\BLam^{(N)}_L(\uu ))$.

\begin{Def} Fix $E\in \R$ and $m>0$. An $N$-particle box
$\BLam^{(N)}_L(\uu)$ is said to be $(E,m)$-non-singular
(in short: $(E,m)${\rm{-NS}}) if the GFs
$\uG^{(N)}_{\BLam^{(N)}_L(\uu)}(E;\uu,\uu')$ defined by
{\rm{(1.15)}} for the Hamiltonian
$\uH^{(N)}_{\BLam^{(N)}_L(\uu)}$ from {\rm{(1.8)}} satisfy
$$\truc{\max}{}{\uy\in\pt \BLam^{(N)}_L(\uu)}
\left| \uG^{(N)}_{\BLam^{(N)}_L(\uu)}(E;\uu,\uy) \right| \leq e^{-m L}.
\eqno(1.16)
$$
\psn
Otherwise, it is called $(E,m)$-singular (or $(E,m)${\rm{-S}}).

A similar concept can be introduced for any finite
set $\BLam^{(N)}\subset\Z^{Nd}$.
\end{Def}

\begin{Def} Let $n$ be a positive integer and $\cJ$ be a non-empty subset of $\{1,\ldots , n\}$.
We say that box $\BLam^{(n)}_L(\uy)$ is $\cJ$-separable from a box
$\BLam^{(n)}_L(\ux)$ (or, equivalently, a point $\uy\in\Z^d$ is
called $\cJ,L$-separable from a point $\ux$) if
$$
\left( \bigcup_{j\in \cJ} \Pi_j \BLam^{(n)}_L(\uy) \right)\cap
\left( \bigcup_{i \not \in \cJ} \Pi_i \BLam^{(n)}_L(\uy) \;
\cup \Pi \BLam^{(n)}_L(\ux)  \right) = \emptyset.\eqno (1.17)
$$
A pair of boxes $\BLam^{(n)}_L(\ux)$, $\BLam^{(n)}_L(\uy)$ is said to be
separable  (or, equivalently,
a pair of points $\ux, \uy\in \Z^{nd}$ is called $L$-separable) if,
for some $\cJ\subseteq\{1,\ldots , n\}$, either
$\BLam^{(n)}_L(\uy)$ is $\cJ$-separable from a box $\BLam^{(n)}_L(\ux)$, or
$\BLam^{(n)}_L(\ux)$ is $\cJ$-separable from a box $\BLam^{(n)}_L(\uy)$.
\end{Def}

The notion of separability of boxes is designed so as to enable
us to establish Wegner--Stollmann type bounds; see Eqns (2.2), (2.3).

In Lemma \ref{CondGeomSep} we give a geometrical upper bound
for the set of points $\uy$ which are {\it not}
separable from a given point $\ux$.

\begin{Lem}\label{CondGeomSep}
Given an $n\geq 2$, let $\ux \in\Z^{nd}$ be an $n$-particle
configuration. For any $L>1$, there exists a finite collection of $n$-particle
boxes $\BLam_{{\wt L}^{(l)}}({\wt\ux}^{(l)})$,
$l=1, \ldots, K(\ux,n)\leq n^n/n!$,
of sides ${\wt L}^{(l)}\leq 5nL$  such that if
a configuration $\uy\in\Z^{nd}$ satisfies
$$
\uy \not \in \bigcup_{\ell=1}^{K(\ux,n)} {\wt\BLam}^{(l)}
\eqno(1.18)
$$
then the boxes $\BLam^{(n)}_L(\ux)$ and $\BLam^{(n)}_L(\uy)$  are separable.
\end{Lem}

The proof of Lemma \ref{CondGeomSep} is given in Section 6.
\psn
The following Theorem \ref{thmone} is completely analogous to  Theorem 2.3 in \cite{DK}
and to Theorem \ref{thmone} in \cite{CS08}, and so is its proof, which we omit.
The reader can check, by inspecting the proofs
in the single-particle case (\cite{DK}) and in the two-particle
case (\cite{CS08}) that
the only modification which causes concern is the choice
of intermediate constants, depending on $N$. However, the core argument of the proof remains unchanged.

\begin{Thm}\label{thmone}
Let $I\subseteq \R$ be a bounded interval. Assume that for
some $m_0>0$ and $L_0/2>1$,
$\lim\limits_{k\to\infty} m_k  \geq m_0/2$,
and for any $k\geq 0$ the following properties hold:
$$
\DSk \;
\begin{array}{l}\hbox{If two boxes $\BLam^{(N)}_{L_k}(\uu)$, $\BLam^{(N)}_{L_k}(\uv)$ are
separable, then} \\
\pr{\forall \,E\in I:\,\BLam^{(N)}_{L_k}(\uu)\;{\rm{or}}\;\BLam^{(N)}_{L_k}(\uv)\;
{\rm{is}}\;(m_k,E){\rm{-NS}}}\geq 1 - L_k^{-2p(N)}.
\end{array}
\eqno(1.19)
$$
Here $L_k$ and $m_k$ are defined in {\rm{(1.11)}},  {\rm{(1.12)}},
and $\sigma$ by {\rm{(1.4)}},
with $p$, $\alpha$ and $\gamma$ satisfying  {\rm{(1.14)}}. Then, for $|g|$ large enough,
with probability one, the spectrum of operator $\uH^{(N)}(\om)$ in
$I$ is pure point. Furthermore, there exists a constant
$m_+\geq m_0/2$ such that all
eigenfunctions
$\Psi_j(\ux;\om)$ of $\uH^{(N)}(\om)$ with eigenvalues
$E_j(\om)\in I$ decay exponentially fast at infinity,
with the effective mass $m_+$:
$$
|\Psi_j(\ux;\om)|\leq C_j(\om) \, e^{-m_+\|\ux\|}.
\eqno(1.20)
$$
\end{Thm}

In future, the eigenvectors of finite-volume Hamiltonians appearing in arguments and calculations, will be assumed normalised. We stress that it is the property \DSk $\,$ encapsulating decay of the GFs which enables the $N$-particle MSA scheme to work. (Here and below, DS stands for `double singularity').
\psn

Clearly, Theorem 1.1 would be proved, once the validity of property \DSk $\,$ is established for all $k\geq 0$.

Our strategy, as indicated in the title of this paper and mentioned earlier in this section, is an induction on the number of particles $N\ge 1$. The base of this induction had been established earlier, starting from papers \cite{FS}, \cite{FMSS}, \cite{DK}, with the help of the MSA, and also in \cite{AM93}, \cite{ASFH01}, in a different way, with the help of the FMM. This allows us to use results of the single-particle localisation theory. We show in this paper that, assuming a certain number of facts established for systems with $n=1, \ldots, N-1$ particles, one can establish similar facts for $N$-particle systems. Once these facts, mostly concerning the decay properties of Green's functions in finite boxes, are established for $N$-particle systems, they imply, in a fairly standard way (essentially, in the same way as in the single-particle and in the two-particle \cite{CS09} theories) the spectral localization for $N$-particle systems. So, according to this plan, we assume established all necessary properties of $n$-particle systems, $1 \le n \le N-1$, and use them whenever necessary. Of course, these properties have to be re-established for $n=N$.
When appropriate, we discuss technical details of proofs in previous works, where the required properties have been proved for $n=1$.

In other words, our paper is organised as a proof of the induction step from $N-1$ to $N$ particles. Within this induction step, we use another inductive scheme - the MSA - where some properties of Green's functions are proved  first at an initial scale $L_0$, and then recursively derived for $N$-particle boxes of sizes $L_k$, $k\ge 1$.

The main property that we have to verify for a  given $N$ and for all $L_k$, $k\ge 0$, is \dsk{N}. Further, the main technical parameter is the exponent $p = p(N) = p(N,g)$ figuring in the RHS of \dsk{N}.   At the initial step
of induction in $N$, we use an important fact from the single-particle theory \cite{DK}: one can guarantee any (arbitrarily large) value $p(1,g)$, provided that $|g|$ is large enough. Cf. (1.14).  Then we show that a similar property holds for any $N$ and for $k=0$, i.e., for the scale $L_0$ (cf. Theorem \ref{MSAInd0}). Therefore, in our double induction scheme (on $N$ and, for a given $N$, on $k$), we require $|g|$ to be sufficiently large so as to guarantee:
\psn
(i) property \dsk{n} for all $k\ge 0$ and for $n=1, \ldots, N-1$ (this property is defined  verbatim, following (1.19) mutatis mutandis);
\psn
(ii) property \dsk{N} for $k=0$.

Parameter $q = q(N) = q(N,g)$ is controlled via Wegner--Stollmann type bounds {\bf (WS1.$n$)}, {\bf (WS2.$n$)}
in (2.2), (2.3), which are proved for all scales $L_k$ at once, without induction in $k$.

\section{The $N$-particle MSA scheme}
\label{InductiveScheme}

In view of Theorem \ref{thmone}, our aim is to check property \DSk $\,$
in Eqn (1.19). We now outline the $N$-particle MSA which is used for this
purpose. In both
single- and $N$-particle versions, the MSA scheme is an elaborate scale induction in $k$ dealing with GFs
$\uG_{\BLam^{(N)}_{L_k}(\uu )} = \uG^{(N)}_{\BLam^{(N)}_{L_k}(\uu )}$ and involving several mutually related parameters;  some of them have been used in Sections 1 and 2.  For a detailed discussion of the role of each parameter,
 see \cite{CS09}.

We will focus in the rest of the paper on the aforementioned \textit{scale} induction in $k$, along sequences $\{(L_k,m_k)\}$
outlined in (1.11), (1.12). Consequently, in some definitions below we refer to the particle number parameter $n\geq 1$, whereas in other definitions - where we want to stress the passage from $N-1$ to $N$ - we will use the capital letter.
\psn

\begin{Def}\label{}
Given $n\ge 1$, $E\in\R$, $\uv\in\Z^{nd}$ and  $L\geq2$, we call the $n$-particle box $\BLam^{(n)}_L(\uv)$ $E$-resonant (briefly: $E${\rm -R}) if the spectrum of the Hamiltonian $H^{(n)}_{\BLam^{(n)}_L(\uv )}$ satisfies
$$
{\rm dist}\left[E, {\rm{spec}}\left(H^{(n)}_{\BLam^{(n)}_{L}(\uv )}
\right)\right]< e^{-L^\beta},
\; \text{ where } \beta = 1/2.
\eqno(2.1)
$$
Box $\BLam^{(n)}_L(\uv)$ is called $E$-completely non-resonant (briefly: $E${\rm -CNR}) if it is $E${\rm -NR} and does not
contain any  $E${\rm -R} box of size $L^{1/\alpha}$.
\end{Def}

Throughout this paper, we use parameter $\beta$ instead  of its value, $1/2$. As with $\alpha = 3/2$, this may be helpful to readers familiar with \cite{DK} and make our notations less cumbersome.

Given $n\ge 1$ and $L_0\geq 2$, introduce the following properties {\bf (WS1.$n$)} and {\bf (WS2.$n$)} of random  Hamiltonians $H^{(n)}_{\BLam^{(n)}_l}$,
$l\geq L_0$.
$$
\hbox{{\bf (WS1.$n$)} $\quad$
$\forall$ $l\geq L_0$, box $\BLam^{(n)}_l(\ux)$ and $E\in\R$: $\;$}
\pr{ \BLam^{(n)}_l(\ux) \text{ is } E\text{\rm-R} }< l^{-q}.
\qquad\;\;\;
\eqno (2.2)
$$

$$
\hbox{\bf (WS2.$n$) $\quad$}\begin{array}{l}\hbox{
$\forall \, l\,\geq L_0$  \hbox{ and separable boxes $\BLam^{(n)}_\ell(\ux)$
and $\BLam^{(n)}_\ell(\uy)$,}}\\ \;
\pr{ \exists \, E\in\R:\;{\rm{both}}\;\BLam^{(n)}_l(\ux) \text{ and }
\BLam^{(n)}_l(\uy) \text{ are } E{\rm{-R}} } < l^{-q}.\end{array}\quad\;\;
\eqno (2.3)
$$
Here $q=q(n)$ is the parameter mentioned in (1.13), (1.14).
\psn

As we already said, the initial step of the $N$-particle MSA scheme consists in establishing
properties \Szero $\,$ and \DSzero; see Eqns (2.4) and (1.19). The
inductive step of the $N$-particle MSA consists in deducing property
\DSkone $\,$  from property \DSk; again see Eqn (1.19).
Both the initial
and the inductive step  will be done with the assistance
of properties {\bf (WS1.$n$)} and/or {\bf (WS2.$n$)}, $n=1$, $\ldots,$ $N$, which have to be
proved independently of the scale induction. In our context, properties {\bf (WS1.$n$)} and {\bf (WS2.$n$)} have been established in \cite{CS08}, Theorems 1, 2. (Despite the fact that properties {\bf (WS1.$n$)}
and {\bf (WS2.$n$)}
had been stated  \cite{CS08} for $n=2$, their proof is automatically
extended to the case of a general $n$.) For reader's convenience
we repeat the corresponding assertion:
\pmn

\begin{Lem}\label{WW}
Under the above assumptions on $\{V(x;\om)\}$ and $U$ (see {\rm (1.3)-(1.5)}), properties
{\rm \textbf{(WS1.$n$)}}, {\rm \textbf{(WS2.$n$)}} $\forall$ positive integer $n$.
\end{Lem}
\psn

Let $I\subseteq\R$ be an interval. Given $m_0>0$ and $L_0\geq 2$, consider property \Szero$\,$:

$$
\Szero \quad \begin{array}{l}\hbox{$\forall$ $\ux\in\Z^{Nd}$,}
\;\;\pr{\exists \, E\in I:\;\;\BLam^{(N)}_{L_0}(\ux)
\;{\rm{is}}\;(E,m_0){\rm -S} }<L_0^{-2p}. \quad\quad\quad
\end{array}
\eqno (2.4)
$$
Here $p=p(N)$ is the parameter mentioned in (1.13), (1.14).

The initial MSA step is summarised in the Theorem \ref{MSAInd0} below. It is completely analogous to
Proposition A.1.2 in \cite{DK}, and so is its proof. Note as well that multi-particle analogs of Propositions A.1.1 and A.1.3 from \cite{DK} can also be proved in the same way as in \cite{DK}. The reason for that is that the multi-particle structure of the external potential $W(\ux;\om)$ and the presence of a bounded interaction potential  $U(\ux)$ (as well as the form of $U(\ux)$ in (1.4)) are  virtually irrelevant for these statements.

\begin{Thm}\label{MSAInd0} $\forall$ given $m_0$ and $L_0\geq 2$ and $\forall$ bounded interval
$I\subset\R$, there exists $g^*_0=g^*_0(N, m_0,L_0,I)\in (0,+\infty)$ such that for
$|g|\geq g^*_0$:
\psn
{\rm (A)} Properties \Szero $\,$ and \DSzero $\,$ hold true.
\psn
{\rm (B)} Moreover, there exists a function
$\widetilde{g}:\,\widetilde{p}\in (d,+\infty)  \mapsto \widetilde{g}(\widetilde{p})\in [g^*_0,+\infty)$
such that if $|g|\ge \widetilde{g}(\widetilde{p})$, then
Eqn {\rm (2.4)} is satisfied with $p = \widetilde{p}$. Equivalently, there exists a function $p(N,g)$ of parameter
$g\in [g^*_0,+\infty)$ (referred to in {\rm (1.13), (1.14)}) such that $p(N,g)\to\infty$ as $|g|\to\infty$ and Eqn {\rm (2.4)} is satisfied with $p=p(N,g)$.
\end{Thm}

To complete the inductive MSA step, we will prove

\begin{Thm}\label{MSAInd} $\forall$ given $m_0>0$, there exist
$g^*_1\in (0,+\infty)$ and $L^*_1\in (0,+\infty )$ such that
the following statement holds. Suppose that $|g|\geq g^*_1$ and
$L_0\geq L^*_1$. Then, $\forall$ $k=0,1,\ldots$ and $\forall$
interval $I\subseteq\R$, property
\DSk $\,$ implies \DSkone.
\end{Thm}

The proof of Theorem \ref{MSAInd} occupies the rest of the paper.
Before we proceed further, let us repeat that the property
\DSk $\,$ for $\forall$ $k\geq 0$ and $\forall$ unit interval $I\subset\R$,
follows directly from Theorems \ref{MSAInd0} and \ref{MSAInd}.

To deduce property \DSkone $\,$  from \DSk, we introduce

\begin{Def}\label{DefFIPI}
Given $R>0$, consider the following set in $\Z^{Nd}$:
$$
\cD_{R} = \big\{\ux=(x_1, \ldots, x_N)\in\Z^{Nd}:\;\max_{1\leq j_1, \,j_2 \leq N}  \|x_{j_1} - x_{j_2}\| \leq NR
\big\}
\eqno(2.5)
$$
It is plain that, with $R=r_0$,  if $\ux\in\cD_{r_0}$ then there is no subset
$\cJ\subset\{1,\ldots ,N\}$ with $1\leq \card\;\cJ<N$ and
$$
\min_{j_1\in\cJ, \,j_2\not\in\cJ} \|x_{j_1} - x_{j_2}\|> r_0.
$$
An $N$-particle box $\BLam^{(N)}_L(\uu)$ is called {\rm fully interactive} when
$\BLam^{(N)}_L(\uu) \cap \cD_{r_0} \neq \emptyset$, and {\rm partially interactive}
if $\BLam^{(N)}_L(\uu) \cap \cD_{r_0}=\emptyset$. For brevity, we use the terms an {\rm{FI-}}box
and a {\rm{PI-}}box, respectively.
\end{Def}

The procedure of deducing property \DSkone $\,$ from \DSk $\,$
is done here separately for the following three cases.
\psn
(I) Both $\BLam^{(N)}_{L_{k+1}}(\ux)$ and $\BLam^{(N)}_{L_{k+1}}(\uy)$ are
PI-boxes.
\psn
(II) Both $\BLam^{(N)}_{L_{k+1}}(\ux)$ and $\BLam^{(N)}_{L_{k+1}}(\uy)$ are FI-boxes.
\psn
(III) One of the boxes is FI, while the other is PI.
\psn

These three cases are  treated in Sections \ref{Case_I}, \ref{Case_II} and \ref{Case_III}, respectively.
The end of Section 5 will mark the end of the proof of Theorem \ref{MSAInd}.
 We repeat
that all cases require the use of property {\bf (WS1.$N$)} and/or {\bf (WS2.$N$)}.

\section{Case I: Partially interactive pairs of singular boxes}
\label{Case_I}

In this section, we aim to derive property \DSkone $\,$
for a pair of partially interactive  and separable boxes
$\BLam^{(N)}_{L_{k+1}}(\ux)$, $\BLam^{(N)}_{L_{k+1}}(\uy)$. Recall, we
are allowed to assume property \DSk $\,$
for every pair of separable boxes $\BLam^{(N)}_{L_k}(\widetilde{\ux})$,
$\BLam^{(N)}_{L_k}(\widetilde{\uy})$, where $\ux,\uy,\widetilde{\ux},
\widetilde{\uy}\in \Z^{Nd}$.
In fact, we will be able to establish property \DSkone $\,$
for partially interactive  separable boxes $\BLam^{(N)}_{L_{k+1}}(\ux)$, $\BLam^{(N)}_{L_{k+1}}(\uy)$
directly, without referring to \DSk. (However, in cases (II)
and (III) such a reference will be needed.)

Let $\BLam^{(N)}_{L_{k+1}}(\uu)$ be an PI-box and write
$\uu =(u_1,\ldots ,u_N)$ as a pair
$(\uu', \uu'')$ where $\cJ$ is a non-empty
subset of $\{1,\ldots ,N\}$ figuring in Definition \ref{DefFIPI}, and
$\uu'=\uu_{\cJ}\in\Z^{\cJ}$ and $\uu''
=\uu_{\cJ^{\rm c}}\in\Z^{\cJ^{\rm c}}$ are the corresponding
sub-configurations in $\uu$: $\uu'
=(u_j,\;j\in\cJ)$ and $\uu''=(u_j,\;j\not\in\cJ)$. Set: $n'=\card\;\cJ$
and $n''=N-n'$. It is convenient to represent $\BLam^{(N)}_L(\uu )$ as
the Cartesian product
$$
\BLam^{(N)}_{L_{k+1}}(\uu) = \BLam_{L_{k+1}}^{(n')}(\uu') \times
\BLam_{L_{k+1}}^{(n'')}(\uu'')
$$
and write $\ux=(\ux',\ux'')$ in the same fashion as $(\uu',\uu'')$.
Correspondingly,
the Hamiltonian $\uH^{(N)}_{\BLam^{(N)}_{L_{k+1}}(\uu)}$ can be written
in the form
$$
\uH \Bphi (\ux)
=\sum\limits_{\uy\in\BLam^{(N)}_{L_{k+1}}(\uu):\atop{\|\uy - \ux\|=1}}
\,\Bphi(\uy)+ \big[U(\ux')+gW(\ux';\om)+U(\ux'')+gW(\ux'';\om)\big]\Bphi(\ux),\\
\eqno(3.1)
$$
or, algebraically,
$$
H^{(N)}_{\BLam^{(N)}_{L_{k+1}}(\uu)} = H^{(n')}_{1;\BLam^{(N)}_{L_{k+1}(\uu')}}
\otimes {\mathbf I} + {\mathbf I}
\otimes H^{(n'')}_{2;\BLam^{(N)}_{L_{k+1}(\uu'')}}.
\eqno (3.2)
$$
Here ${\mathbf I}$ is the identity operator on the complementary
variable.

Due to the symmetry of terms $U$ and $W$, in the forthcoming
argument we can assume, without loss of generality, that
$$\cJ =\{1,\ldots ,n'\},\;\;\cJ^{\rm c}=\{n'+1,\ldots, N\}.$$

\begin{Def}\label{DefBad}
Let be $n\in \{1, \ldots, N-1\}$, $k\ge 0$ and $\uu'=(u_1, \ldots, u_{n})
\in\Z^{nd}$. Given a bounded interval $I\subset \R$ and $m>0$, the $n$-particle box
$\BLam^{(n)}_{L_k}(\uu')$ is called $m$-tunneling
($m${\rm{-T}}, for short) if
$
\exists\, E\in I$  and disjoint $n$-particle boxes $\BLam^{(n)}_{L_{k-1}}(\uv_1),
\Lam^{(n)}_{L_{k-1}}(\uv_2) \subset \BLam^{(n)}_{L_{k}}(\uu')$  which are $(E,m)${\rm-S}.
An $N$-particle box of the form
$\BLam^{(N)}_{L_k}(\uu) = \Lam^{(n')}_{L_{k-1}}( \uu') \times \Lam^{(n'')}_{L_{k-1}}( \uu'')$,
with $n'+n''=N$, $\uu = (\uu',\uu'')$, $\uu' =(u_1, \ldots, u_{n'})$, $\uu'' =(u_{n'+1}, \ldots, u_{N})$,
is called $(m,n',n'')${\rm{-NT}} if both
$\Lam^{(n'}_{L_{k-1}}( \uu')$ and $\Lam^{(n'')}_{L_{k-1}}( \uu'')$ are $m${\rm-NT}. Otherwise, it is
called $(m,n',n'')${\rm-T}. Finally, a box $\BLam^{(N)}_{L_k}(\uu)$ is called $m${\rm-T}
if it is $(m,n',n'')${\rm-T} for some $n', n''\geq 1$ with $n'+n''=N$, and $m${\rm-NT},
otherwise.
\end{Def}

The following statement will be sometimes referred to as the \textbf{ NITRoNS} property of PI-boxes:
\textit{Non-Interacting boxes are Tunneling, Resonant or (otherwise) Non-Singular}.
Cf. \cite{CS09}.

\begin{Lem}\label{GF_rep}
Consider an $N$-particle  box
$
\BLam^{(N)}_{L_k}(\uu)$ of the form $\BLam^{(n')}_{L_{k-1}}( \uu') \times \BLam^{(n'')}_{L_{k-1}}( \uu''),
$
where
$\uu = (\uu',\uu''), \; \uu' =(u_1, \ldots, u_{n'})\in\Z^{n'd}, \; \uu'' =(u_{n'+1}, \ldots, u_{N})\in\Z^{n''d}.$
Assume that $\forall\,j_1, j_2$ with $1 \leq j_1 \leq n'$, $n'+1 \leq j_2 \leq N$, we have $\|u_{j_1} - u_{j_2}\|>r_0$,
so that $\BLam^{(N)}_{L_k}(\uu)$ is {\rm PI}.
Assume also that $\BLam^{(N)}_{L_k}(\uu)$ is $E${\rm-CNR} and $m${\rm-NT}. Let
$$
\left\{ (\lam_a, \ffi_a), \; a = 1, \ldots, |\BLam^{(n')}_{L_{k}}( \uu')|\right\},\;\;
\left\{ (\mu_b, \psi_b), \; b = 1, \ldots, |\BLam^{(n'')}_{L_{k}}( \uu'')| \right\},
$$
be the eigenvalues and eigenvectors of  $\uH^{(n')}_{\BLam^{(n')}_{L_{k}}( \uu')}$ and
$\uH^{(n'')}_{\BLam^{(n'')}_{L_{k}}( \uu'')}$, respectively. Set
$$
m' = m\left(1 - L_k^{-(1-\beta)} - L_k^{-1} \ln L_k^{N(d-1)}  \right).
$$
Then we have
$$
\max_{1 \leq a \leq |\BLam^{(n')}_{L_{k}}( \uu')| }\,
\max_{\uv''\in\pt \BLam^{(n'')}_{L_k}(\uu'')} \,| G^{(n'')}(\uu'', \uv''; E - \lam_a)|
\leq e^{-m'L_k}
$$
and, similarly,
$$
\max_{1 \leq b \leq |\BLam^{(n'')}_{L_{k}}( \uu'')|}\,
\max_{\uv'\in\pt \BLam^{(n')}_{L_k}(\uu')} \,| G^{(n')}(\uu', \uv'; E - \mu_b)|
\leq e^{-m'L_k}.
$$
As a consequence, the $N$-particle box  $\BLam^{(N)}_{L_k}(\uu)$ is $(E,m')${\rm-NS}.
\end{Lem}
The proof of Lemma \ref{GF_rep} is given in Section 7. (It is fairly straightforward and based on the representations (7.1) - (7.3).)
\pn
\begin{Lem}\label{lem_one}
Let $n,k$ be positive integers and suppose that \dsk{n} holds true. Then
$$
\pr{\BLam^{(n)}_{L_k}(\uy) \text{ is } m\text{\rm -T}} \leq \half |\BLam^{(n)}_{L_k(\uy)}|^2 \, L_{k-1}^{-2 p(n)}
= \half L_k^{-\frac{2p(n)}{\alpha} + 2d}.
\eqno(3.3)
$$
Here $p(n)$ is the parameter figuring in {\rm (1.13), (1.14)}.
\end{Lem}
\proof  Combine \dsk{n} with a straightforward (albeit not sharp) upper bound
$\half |\BLam^{(n)}_{L_k(\uy)}|^2$ for the number of pairs of centers
$\uv_1, \uv_2$ of boxes
$\BLam^{(n)}_{L_{k-1}}(\uv_1), \BLam^{(n)}_{L_{k-1}}(\uv_2) \subset \BLam^{(n)}_{L_{k}}(\uy)$. \qed

In Lemma \ref{lemoneB} we assume for simplicity that a PI box $\BLam^{(N)}_{L_k}(\uy)$
corresponds to an $N$-particle system that splits into two subsystems, with particles $1, \ldots, n'$
and $n'+1, \ldots, n'+n''=N$, respectively, and the two subsystems do not interact with each other.

\begin{Lem}\label{lemoneB}
Let $\BLam^{(N)}_{L_k}(\uy)$ be an $N$-particle {\rm PI} box, with
$$
\BLam^{(N)}_{L_k}(\uy) = \BLam^{(n')}_{L_k}(\uy') \times \BLam^{(n'')}_{L_k}(\uy''),
$$
where $n',n'' \geq 1$, $n' + n''=N$; $\uy = (\uy', \uy'')$, $\uy' = (y_1, \ldots y_{n'}) \in \Z^{n'd}$,
$\uy'' = (y_{n'+1}, \ldots y_{N}) \in \Z^{n''d}$, and
$$
\min_{1 \le i \le n' } \;\; \min_{n'+1 \le j \le N } \| y_i - y_j\| > r_0.
$$
Then for any given value $p(N)>0$ there exists $g^*_2\in(0,+\infty)$ such that if $|g|\ge g^*_2$, then
$$
\pr{\BLam^{(N)}_{L_k}(\uy) \text{ is } m\text{\rm -T}} \leq \half L_k^{-2p(N)}.
\eqno(3.4)
$$
\end{Lem}

\proof
By Definition \ref{DefBad}, box $\BLam^{(N)}_{L_k}(\uy)$ is $m$-T iff at least one of constituent boxes
$\BLam^{(n')}_{L_k}(\uy')$, $\BLam^{(n'')}_{L_k}(\uy'')$ is $m$-T. By virtue of Lemma \ref{lem_one},
inequality (3.3) holds for both $n=n'$ and $n=n''$. This leads to the assertion of Lemma \ref{lemoneB}.
\qed

\psn
\textbf{Remark.} The assertion of Lemma \ref{lemoneB} remains true for a general type of interaction
(with appropriate modifications), but is simpler and more transparent in the case of two-body interaction of the form (1.4). This explains our choice of the interaction energy function $U(\ux)$. Besides, in applications to the electron transport problems, such a choice is perfectly justified: here, a commonly accepted form of interaction
is two-body Coulomb.

\psn

We repeat that, according to the structure of the MSA scheme, for any given number of particles $n=1, \ldots, N$, any (i.e., arbitrarily large) values $p(n)$,
$q(n)$ can be used, provided that $|g|$ is sufficiently large. In other words, parameters
$p(n), q(n)$ follow (1.14). Indeed, for $p(n)$ this can be guaranteed, by direct inspection,
 for the boxes of initial size $L_0$. Cf. Appendix in \cite{DK}. The same property is  then reproduced inductively at any scale $L_k$,
$k\geq 1$. As to $q(n)$, one can actually obtain a stronger bound:
$$
\pr{ \BLam^{(N)}_{L_k}(\uu) \text{ is } E\text{-R} } \le e^{-L_k^\beta } \ll L_k^{-s}
$$
for any a priori given $s$ including $s=q(N)$, provided that $\beta>0$ and $L_0$ (hence, any $L_k$) is large enough.
\pmn

\begin{Lem}\label{TwoOffDiagSing}
 Assume that property {\bf (WS2.$N$)} and Eqns {\rm (3.3), (3.4)} hold true.  Suppose also that
$|g|$ is sufficiently large, so that for all $n=1, \ldots, N-1$ the bound (3.3) holds
with $p(n) \ge 2p(N) + 2d$, and that $L_0$ is sufficiently large, so that for any
$k\ge 0$ we have
$$
L_k^{-\frac{2p(n)}{\alpha} + 2d} \leq \frac{1}{4} L_k^{-2p(N)}.
$$
Then, $\forall$ interval $I\subseteq\R$, $\forall$ integer
$k\geq 0$ and $\forall$ pair of {\rm separable} {\rm PI}
$N$-particle boxes $\BLam^{(N)}_{L_{k}}(\ux)$ and
$\BLam^{(N)}_{L_{k}}(\uy)$,
$$
\pr{\exists\;E\in I:\, \BLam^{(N)}_{L_{k}}(\ux)
\text{ \rm and } \BLam^{(N)}_{L_{k}}(\uy)
\text{ \rm are } (E,m_{k}){\rm -S} }
\leq  \frac{1}{2} L_k^{-2p(N)} + L_{k}^{-q(N)}.
\eqno (3.5)
$$
Here $p(N), q(N)$ are the parameters from {\rm (1.13)}.
\end{Lem}

\myproof{Lemma {\rm{\ref{TwoOffDiagSing}}}} By virtue of Lemma 3.1,
$$
\begin{array}{l}
\pr{\exists\;E\in I:\, \BLam^{(N)}_{L_{k}}(\ux)
\text{ \rm and } \BLam^{(N)}_{L_{k}}(\uy)
\text{ \rm are } (E,m_{k}){\rm -S} }\\
 \leq
\pr{\BLam^{(N)}_{L_{k}}(\ux) \text{ or } \BLam^{(N)}_{L_{k}}(\uy) \text{ is } m_k{\rm -T}} \\
+ \;\pr{\exists\;E\in I:\, \BLam^{(N)}_{L_{k}}(\ux)
\text{ \rm and } \BLam^{(N)}_{L_{k}}(\uy)
\text{ \rm are } E{\rm -R} }.
\end{array}
\eqno(3.6)
$$
Observe that, by \textbf{NITRoNS}, if $\BLam^{(N)}_{L_{k}}(\uy)$ is both $E$-NR and  $m_k$-T,  and it has the form
$$
\BLam^{(N)}_{L_{k+1}}(\uu) = \BLam_{L_{k+1}}^{(n')}(\uu') \times
\BLam_{L_{k+1}}^{(n'')}(\uu'')
$$
with  $\BLam_{L_{k+1}}^{(n')}(\uu')$ and
$\BLam_{L_{k+1}}^{(n'')}(\uu'')$ not interacting with each other, so that Eqn (3.2) holds,
then at least one of these "projection" boxes must be $m_k$-T. Without loss of generality, assume that $\BLam_{L_{k+1}}^{(n')}(\uu')$ is $m_k$-T  and set $\elli=n'$. A similar argument applies, of course, to the case where  $\BLam_{L_{k+1}}^{(n'')}(\uu'')$ is $m_k$-T. Naturally, $\ell \le N-1$, so that by the hypothesis of the lemma and by Eqns (3.3), (3.4) we have
$$
\begin{array}{l}
\pr{\BLam^{(N)}_{L_{k}}(\ux) \text{ or } \BLam^{(N)}_{L_{k}}(\uy) \text{ is } m_k{\rm -T}}
\le 2\cdot \frac{1}{4}L_k^{-\frac{2p(\elli)}{\alpha} + 2d}
\le L_k^{-2p(N)} /2.
\end{array}
\eqno(3.7)
$$
Now the assertion of Lemma \ref{TwoOffDiagSing} follows from Eqn (3.6) and (3.7). \qquad\qed
\pmn
\textbf{Remark. } It is readily seen that the RHS of Eqn (3.5) is bounded by $L_k^{-2p(N)}$,
provided that $L_k^{-q(N)} < L_k^{-2p(N)}/2$, i.e., for $q(N)$ large enough.
\psn

An immediate corollary of Lemma \ref{TwoOffDiagSing} is the following
\psn
\begin{Thm} $\forall$ given interval $I\subseteq\R$
and $k=0,1,\ldots$, property \DSk $\,$
holds for all pairs of separable {\rm PI}-boxes
$\BLam^{(N)}_{L_k}(\ux)$, $\BLam^{(N)}_{L_k}(\uy)$.
\end{Thm}

Summarising the above argument: as was said earlier, verifying property \DSkone $\,$
for a pair of $N$-particle
PI-boxes did not force us to assume \DSk.
However, in the course of deriving \DSkone $\,$
for PI-boxes we used
property {\bf (WS2.$N$)}.

This completes the analysis of the case (I) where both boxes
$\BLam^{(N)}_{L_{k+1}}(\ux)$ and $\BLam^{(N)}_{L_{k+1}}(\uy)$ are PI.

For future use, we also give

\begin{Lem}\label{LemonM}
Consider a $N$-particle box $\BLam^{(N)}_{L_{k+1}}(\uu)$.
Let $M=M(\BLam^{(N)}_{L_{k+1}}(\uu);E)$ be the maximal number of $(E,m_k)${\rm -S},
pair-wise {\rm{separable}}$\;$
{\rm{PI-}}boxes $\BLam^{(N)}_{L_k}(\uu^{(l)})\subset \Lam_{L_{k+1}}(\uu)$.
The following property holds
$$
\pr{\exists E\in I:\;\; M(\BLam^{(N)}_{L_{k+1}}(\uu);E) \geq 2 }
\leq  L_k^{2d\, \alpha} \cdot \left( \frac{1}{2} L_k^{-2p'(N-1)} + L_{k}^{-q(N)} \right),
\eqno (3.8)
$$
where
$$
p'(N-1,g) := \min\{ p(n,g), \, 1 \le n \le N-1 \} \truc{}{\longrightarrow}{|g|\to \infty+\infty} +\infty.
\eqno(3.9)
$$
\psn
As before, $p(N), q(N)$ are the parameters from in {\rm (1.13), (1.14)}.
\end{Lem}

\begin{proof} The number of possible pairs
of centres $(\uu^{(l_1)}, \uu^{(l_2)})$, $1 \leq l_1 < l_2 \leq M$, is bounded by $L_{k+1}^{2d}$,
while for a given pair of centres one can apply Lemma \ref{TwoOffDiagSing}.
This leads to the assertion of Lemma \ref{LemonM}. \qed
\end{proof}

\section{Fully interactive pairs of singular boxes}
\label{Case_II}

The main outcome in case (II) is Theorem \ref{ThmTwoISing} placed at
the end of this section. Before we proceed further, let us state
a geometric assertion (see Lemma \ref{DistDiag} below)
which we prove in Section 6.

\begin{Lem}\label{DistDiag}
Let be $n\geq 1$, $L> r_0$ and consider two separable $n$-particle {\rm FI}-boxes $\BLam^{(n)}_L(\uu')$ and
$\BLam^{(n)}_L(\uu'')$, with
$\dist\left[\BLam^{(n)}_L(\uu'),\BLam^{(n)}_L(\uu'')\right] >8L$. Then
$$
\Pi\BLam^{(n)}_L(\uu') \cap \Pi\BLam^{(n)}_L(\uu'')= \emptyset.
\eqno (4.1)
$$
\end{Lem}
\pmn

Lemma \ref{DistDiag} is used in the proof of Lemma \ref{ProbISing}
which, in turn, is important in establishing
Theorem \ref{ThmTwoISing}. In fact, Lemma \ref{DistDiag} is a natural development of
Lemma 2.2 in \cite{CS08}.
Let $I\subseteq\R$ be an interval. Consider the following assertion
$$\ISk:
\begin{array}{l}\hbox{$\forall$ pair of  interactive {\rm{separable}} boxes
$\BLam^{(N)}_{L_k}(\ux )$ and $\BLam^{(N)}_{L_k}(\uy )$:}\\
\P\;\Big\{ \exists \, E\in I:\,{\rm{both}}\;
\BLam^{(N)}_{L_k}(\ux ),\;\BLam^{(N)}_{L_k}(\uy )\;{\rm{are}}
\;(E,m_k)\text{-S}\Big\}\leq L_k^{-2p(N)},\end{array}
\eqno (4.2)
$$
with $p(N)$ as in {\rm (1.13), (1.14)}.

\begin{Lem}\label{ProbISing}
Given $k\geq 0$, assume that property \ISk$\,$  holds true. Consider a box
$\BLam^{(N)}_{L_{k+1}}(\uu )$ and let $\widetilde{N}(\BLam^{(N)}_{L_{k+1}}(\uu );E)$ be the maximal number of
$(E,m_k)${\rm -S}, pair-wise {\rm{separable}}\; {\rm{FI-}}boxes
$\BLam^{(N)}_{L_k}(\uu^{(j)})\subset\BLam^{(N)}_{L_{k+1}}(\uu )$.
Then $\forall$ $\elli\geq 1$,
$$
\pr{\exists\;E\in I:\;\; \widetilde{N}(\BLam^{(N)}_{L_{k+1}}(\uu );E) \geq 2\elli }\leq
L_k^{2\elli(1+d\alpha)} \cdot  L_k^{-2\elli p(N)}.
\eqno (4.3)
$$
\end{Lem}

\myproof{Lemma {\rm{\ref{ProbISing}}}}Suppose
$\exists$ FI-boxes $\BLam^{(N)}_{L_k}(\uu^{(1)}),\ldots$,
$\BLam^{(N)}_{L_k}(\uu^{(2n)})$\\ $\subset\BLam^{(N)}_{L_{k+1}}(\uu )$ such that
any two of them are separable.  By virtue of Lemma \ref{DistDiag}, it is readily seen that

(a) $\forall$ pair $\BLam^{(N)}_{L_k}(\uu^{(2i-1)})$, $\BLam^{(N)}_{L_k}(\uu^{(2i)})$,
the respective (random) operators\\
$H^{(N)}_{\BLam^{(N)}_{L_k}(\uu^{(2i-1)})}(\om )$
and $H^{(N)}_{\BLam^{(N)}_{L_k}(\uu^{(2i)})}(\om )$ are mutually independent,
and so are their
spectra and Green's functions $\uG^{(N)}_{\BLam^{(N)}_{L_k}(\uu^{(2i-1)})}$
and $\uG^{(N)}_{\BLam^{(N)}_{L_k}(\uu^{(2i)})}$.

(b) Moreover, the following pairs of operators form an independent family:
$$
\left(\uH^{(N)}_{\BLam^{(N)}_{L_k}(\uu^{(2i-1)})}(\om ),
H^{(N)}_{\BLam^{(N)}_{L_k}(\uu^{(2i)})}(\om )\right),\;\;i=1, \dots, \ell,
\eqno(4.4)
$$

Indeed, operator
$\uH^{(N)}_{\BLam^{(N)}_{L_k}(\uu^{(i)})}$, with $i\in\{1,\ldots, 2n\}$, is
measurable relative to the sigma-algebra
$\ccB(\BLam^{(N)}_{L_k}(\uu^{(i)})$ generated by $\{V(x), \, x\in \Pi \, \BLam^{(N)}_{L_k}(\uu^{(i)}) \}$,
$i=1, \ldots, 2\ell$.
Now, by Lemma 4.2, the sets $\Pi \, \BLam^{(N)}_{L_k}(\uu^{(i)})$,
$i\in\{1,\ldots , 2\elli\}$, are pairwise disjoint,
so that all sigma-algebras $\ccB(\BLam^{(N)}_{L_k}(\uu^{(i)})$, $i\in\{1,\ldots ,2\elli\}$, are
independent.
\psn

Thus, any collection of events $\cA_1$,
$\ldots$, $\cA_{\ell}$ related to the corresponding pairs
$$
\left(\uH^{(N)}_{\BLam^{(N)}_{L_k}(\uu^{(2i-1)})}, \uH^{(N)}_{\BLam^{(N)}_{L_k}(\uu^{(2i)})} \right),
i=1, \ldots, \elli,
$$
also form an independent family. Now, for $i=1,\ldots, \elli-1$,  set
$$
\cA_i = \myset{ \exists \, E\in I:\,
\BLam^{(N)}_{L_k}(\uu^{(2i+1)}) \text{ and } \;\BLam^{(N)}_{L_k}(\uu^{(2i+2)})\;{\rm{are}}\;
(E,m_k)\text{-S} }.
\eqno (4.5)
$$
Then, by virtue of \ISk (see (4.3)),
$$
\P\;\Big\{ \cA_j\Big\}\leq L_k^{-2p(N)},\;\;0\leq j\leq \elli-1,
\eqno (4.6)
$$
and by virtue of independence of events $\cA_0$,
$\ldots$, $\cA_{n-1}$, we obtain
$$
\P\;\Big\{ \bigcap_{j=0}^{\elli-1} \cA_j\Big\} =
\prod_{j=0}^{\elli-1}\P\;\Big\{\cA_j\Big\}\leq
\left(L_k^{-2p(N)}\right)^{\elli}.
\eqno (4.7)
$$
To complete the proof, note that the
total number of different families
of $2\elli$ boxes $\BLam^{(N)}_{L_k}\subset\BLam^{(N)}_{L_{k+1}}(\uu )$
with required properties is bounded from above by
$$
\frac{1}{(2\elli)!} \left[ 2 (L_k/2 + r_0 + 1) L_{k+1}^{d}\right]^{2\elli}
\leq \frac{1}{(2\elli)!} \left(2 L_k L_{k+1}^{d} \right)^{2\elli}
\leq L_k^{2\elli(1+d\alpha)},
$$
since their centres must belong to the subset
$\cD_{L_k+r_0}\cap \BLam^{(N)}_{L_{k+1}}(\uu )$ (see (2.5)).
Recall also that $r_0 < L_0\leq L_k$ $\forall\,k\geq 0$, by our assumption and
by construction. This yields Lemma \ref{ProbISing}.
$\qquad\qed$

\begin{Lem}\label{HowManySub} Let $K(\uu, L_{k+1};E)$ be the maximal
number of $(E,m_k)${\rm -S}, pair-wise {\rm{separable}}
boxes $\BLam^{(N)}_{L_k}(\uu^{(j)})\subset \BLam^{(N)}_{L_{k+1}}(\uu )$ (fully or partially interactive).
Then $\forall$ $\ell\geq 1$,
$$
\pr{\exists E\in I:\;\; K(\uu, L_{k+1};E) \geq 2\ell +2}\leq
L_k^{4d\alpha} \cdot  L_k^{-2 p(N-1)} +
L_k^{2\elli(1+d\alpha)} \cdot  L_k^{-2\elli p(N)},
\eqno (4.8)
$$
where $p(N-1)$ and $p(N)$ are parameters from {\rm (1.13), (1.14)}, for the system with $N-1$ and $N$
particles, respectively.
\end{Lem}
\myproof{Lemma {\rm{\ref{HowManySub}}}}
Assume that $K(\uu, L_{k+1};E) \geq 2\elli+2$.
Let $M(\BLam^{(N)}_{L_{k+1}(\uu )};E)$ be as in Lemma \ref{LemonM}
and $N(\BLam^{(N)}_{L_{k+1}(\uu )};E)$ as in Lemma \ref{ProbISing}. Obviously,
$$
K(\uu, L_{k+1};E) \leq M(\BLam^{(N)}_{L_{k+1}(\uu )};E)
+ N(\BLam^{(N)}_{L_{k+1}(\uu )};E).
$$
Then either $M(\BLam^{(N)}_{L_{k+1}(\uu )};E)\geq 2$ or
$N(\BLam^{(N)}_{L_{k+1}(\uu )};E)\geq 2\elli$. Therefore,
$$\begin{array}{l}
\pr{\exists E\in I:\;\; K(\uu, L_{k+1};E) \geq 2\elli +2} \\
\leq \pr{\exists E\in I:\;\; M(\BLam^{(N)}_{L_{k+1}(\uu )};E) \geq 2}
+ \pr{\exists E\in I:\;\; N(\BLam^{(N)}_{L_{k+1}(\uu )};E) \geq 2\elli}\\
\leq L_k^{4d\alpha} \cdot  L_k^{-2 p(N-1)} + L_k^{2\elli(1+d\alpha)} \cdot  L_k^{-2\elli p(N)},
\end{array}
$$
by virtue of (3.8) and (4.3)
\qed

An elementary calculation now gives rise to the following
\pmn

\begin{Cor}\label{CorHowManySub}  Under assumptions of Lemma
{\rm{\ref{HowManySub}}}, with $\elli\geq 4$, $p(N-1)$ and $p(N)$ large enough and for $L_0$ large enough,
we have, $\forall\,$ integer $k \geq 0$,
$$
\pr{\exists E\in I:\;\; K(\uu, L_{k+1};E) \geq 2\elli +2}\leq
\,   L_{k+1}^{-2p(N)-1}.
\eqno (4.9)
$$
\end{Cor}

Now the Wegner--Stollmann bound {\bf (WS2.$N$)} implies
\begin{Lem}\label{LemCNR} If $N$-particle boxes $\BLam^{(N)}_{L_{k+1}}(\uu')$, $\BLam^{(N)}_{L_{k+1}}(\uu'')$ (fully or partially interactive) are separable, then
$\forall\, L_0> (J+1)^2$,
$$
\begin{array}{r}
\pr{\forall\, E\in I:\, \text{ either } \BLam^{(N)}_{L_{k+1}}(\uu')
 \text { or } \BLam^{(N)}_{L_{k+1}}(\uu'') \text{ is } (E,J){\rm-CNR}}
\qquad\\
\geq 1 - (J+1)^2 L_{k+1}^{-(q(N)\alpha^{-1}-2\alpha)}
> 1 - L_{k+1}^{-(q'(N)-4)}.
\end{array}
\eqno (4.10)
$$
Here $q(N)$ is the parameter from {\rm (1.13)} and $q'(N):= q(N)/\alpha.$
\end{Lem}
\pmn

The statement of Lemma \ref{J_NS} below
is a simple reformulation of Lemma 4.2
from \cite{DK}, adapted to our notations.
Indeed, the reader familiar with
the proof given in \cite{DK} can see that the structure
of the external potential is irrelevant to
this completely deterministic statement. So it applies
directly to our model with potential energy
$U(\ux) + gW(\ux;\om)$. For that reason, the proof of
 Lemma \ref{J_NS} is omitted.

\begin{Lem}\label{J_NS}  Fix an odd positive integer $J$
and suppose that the following properties are fulfilled:
\psn
\centerline{{\rm (i)} $\BLam^{(N)}_{L_{k+1}}(\uv)$ is $(E,J)${\rm{-CNR}}, and
{\rm (ii)} $K(\BLam^{(N)}_{L_{k+1}(\uu )};E)\leq J$.}
\psn

Then for sufficiently large $L_0$, box $\BLam^{(N)}_{L_{k+1}}(\uv)$
is $(E,m_{k+1})${\rm{-NS}} with
$$
m_{k+1} \geq m_k \left( 1 - \frac{5J+6}{L_k^{1/2}} \right)
 > m_0/2>0.
 \eqno (4.11)
$$
\end{Lem}

Now the main result of this section:

\begin{Thm}\label{ThmTwoISing}
Fix a bounded interval $I \subset\R$. For $p(N)$ large enough there exists
$L^*_0\in(0,+\infty)$ such that if $L_0\geq L^*_0$ and $p(N-1)$ is large enough, then, $\forall\,
k\geq 0$, property \ISk  $\,$ in {\rm{(4.2)}} implies \ISkone,  with the same $p(N)$.
\end{Thm}
\pn
{\it Proof of Theorem} \ref{ThmTwoISing}. Let $\ux,\uy\in\Z^{Nd}$
and assume that
$\BLam^{(N)}_{L_{k+1}}(\ux )$ and $\BLam^{(N)}_{L_{k+1}}(\uy )$ are
{\rm{separable}} FI-boxes. Consider the following two events:
$$\cB = \Big\{ \exists \, E\in I:\,{\rm{both}}\;
\BLam^{(N)}_{L_{k+1}}(\ux )
\text{ and } \BLam^{(N)}_{L_{k+1}}(\uy )\;
\text{ are } (E,m_{k+1})\text{-S}\Big\}\,,$$
and, for a given odd integer $J$,
$$\cR = \Big\{\exists \, E\in I:\, \text{ neither }
\BLam^{(N)}_{L_{k+1}}(\ux ) \text{ nor } \BLam^{(N)}_{L_{k+1}}(\uy )
\text{ is } (E,J){\rm{-CNR}} \Big\}.$$
By  virtue of Lemma  \ref{LemCNR}, for $L_0\geq J+1)^2$ and $\alpha = 3/2$, we have:
$$
\pr{\cR} < L_{k+1}^{-(q'(N)-4)}, \; q'(N):= q(N)/\alpha.
\eqno(4.12)
$$
Further,
$\pr{\cB} \leq \pr{\cR} + \pr{\cB\cap\cR^{\rm c}}$,
and we know that $\pr{\cR}\leq L_{k+1}^{-q'(N)+4}$.
So, it suffices now to estimate $\pr{\cB\cap \cR^{\rm c}}$. Within
the event
$\cB\cap \cR^{\rm c}$, for any $E\in I$, either $\BLam^{(N)}_{L_{k+1}}(\ux )$ or
$\BLam^{(N)}_{L_{k+1}}(\uy )$ must be $(E,J)$-CNR. Without loss of generality,
assume that for some $E\in I$, $\BLam^{(N)}_{L_{k+1}}(\ux )$ is
$(E,J)$-CNR and $(E,m_{k+1})$-S. By Lemma
\ref{J_NS}, for such value of $E$, $K(\BLam^{(N)}_{L_{k+1}}(\ux );E) \geq J+1$.
We see that
$$
\cB \cap \cR^{\rm c} \subset \Big\{\exists E\in I:\;\;
K(\BLam^{(N)}_{L_{k+1}}(\ux );E) \geq J+1 \Big\}
$$
and, therefore, by Lemma \ref{HowManySub},
$$
\pr{\cB \cap \cR^{\rm c}} \leq \pr{\exists E\in I:\;\;
K(\BLam^{(N)}_{L_{k+1}}(\ux );E) \geq J+1}
\leq L_k^{-2p(N)}.
\eqno (4.13)
$$
\qquad\qed

{\bf Remark.} The integer $J$ figuring throughout Section 4 depends on $N, d$, and the choice of parameter
$p(N)$. In turn, $p(N)$ is determined by dimension $d$  and the choice of value $\ell$ from Lemma \ref{ProbISing}.
In addition, parameter $p(N-1)$ should be large enough (as was stated in Theorem \ref{ThmTwoISing}).

\section{Mixed pairs of singular $N$-particle boxes}
\label{Case_III}

It remains to derive the property \DSkone $\,$
in case (III),
i.e., for mixed pairs
of $N$-particle boxes (where one is FI and the other PI).
Here we use several properties which have
been established earlier in this paper for all scale lengths,
namely, {\bf(WS1.$n$)}, {\bf(WS2.$n$)} for $n=1, \ldots, N$, \textbf{NITRoNS}, and the inductive assumption \ISkone
which we have already derived from \ISk $\,$ in Section 4.

A natural counterpart of Theorem \ref{ThmTwoISing} for
mixed pairs of boxes is the following

\begin{Thm}\label{ThmINISing} $\forall$ given interval $I\subseteq\R$,
there exists a constant
$L^*_1\in(0,+\infty)$ with the following property. Assume that $L_0\geq L^*_1$ and,
for a given $k\geq 0$, the property \DSk $\,$
holds
{\rm (i)} $\forall$ pair of {\rm separable} $\;${\rm{PI-}}boxes
$\BLam^{(N)}_{L_k}({\wt\ux})$,
$\BLam^{(N)}_{L_k}({\wt\uy})$, and
{\rm (ii)} $\forall$ pair of separable $\;${\rm{FI-}}boxes
$\BLam^{(N)}_{L_k}({\wt\ux})$, $\BLam^{(N)}_{L_k}({\wt\uy})$.

Let $\BLam^{(N)}_{L_{k+1}}(\ux)$,
$\BLam^{(N)}_{L_{k+1}}(\uy)$ be a pair of {\rm separable} boxes, where
$\BLam^{(N)}_{L_{k+1}}(\ux)$ is $\;${\rm FI}$\;$ and
$\BLam^{(N)}_{L_{k+1}}(\uy)\;$ {\rm{PI}}. Then
$$\P\;\Big\{\exists \, E\in I:\,\text{\rm both } \BLam^{(N)}_{L_{k+1}}(\ux),
\;\;
\BLam^{(N)}_{L_{k+1}}(\uy )\;{\rm{are}}\;(E,m_{k+1}){\rm -S}\Big\}
\leq L_{k+1}^{-2p(N)}.\eqno (5.1)$$
\end{Thm}
\pmn


\myproof{Theorem {\rm\ref{ThmINISing}}} Recall that the Hamiltonian
$\uH^{(N)}_{\BLam^{(N)}_{L_{k+1}}(\uy )}$ is decomposed as in Eqns
(3.1), (3.2). Consider the following three events:
$$
\begin{array}{l}
\cB = \Big\{ \exists \, E\in I:\,{\rm{both}}\;
\BLam^{(N)}_{L_{k+1}}(\ux),\;\;\BLam^{(N)}_{L_{k+1}}(\uy)\;
\text{ are } (E,m_{k+1})\text{-S}\Big\}\,, \\
\cT = \Big\{\hbox{  $\Lam_{L_{k+1}}(\uy)$ is $(m_0)$-T} \Big\},\\
\cR = \Big\{\exists \, E\in I:\, \text{ neither }
\BLam^{(N)}_{L_{k+1}}(\ux) \text{ nor }
\BLam^{(N)}_{L_{k+1}}(\uy)\text{ is } (E,J){\text{-CNR}} \Big\}.
\end{array}
$$
Recall that by virtue of (3.4), we have
$$
\pr{ \cT } \le \half L_{k+1}^{-2p(N)}
\eqno (5.2)
$$
For the event $\;\cR\;$ we have, by virtue of Lemma \ref{LemCNR} and inequality (4.13),
$$
\pr{\cR} \leq L_{k+1}^{-q(N)+2};
\eqno (5.3)
$$
as before, $q(N)$ is the parameter from (1.13).
Further,
$\pr{\cB} \leq \pr{T} + \pr{\cB\cap \cT^{\rm c}} \leq
\half L_{k+1}^{-2p(N)} + \pr{\cB\cap \cT^{\rm c}},
$
and we have
$$
\pr{\cB \cap \cT^{\rm c}}
\leq \pr{\cR} + \pr{\cB\cap \cT^{\rm c}\cap \cR^{\rm c}}
\leq L_{k+1}^{-q(N)+2}
+ \pr{\cB\cap \cT^{\rm c}\cap \cR^{\rm c}}.
$$
Within the event
$\cB\cap \cT^{\rm c} \cap \cR^{\rm c}$, either $\BLam^{(N)}_{L_{k+1}}(\ux)$ or $\BLam^{(N)}_{L_{k+1}}(\uy)$
is $E$-CNR. It must be the FI-box $\BLam^{(N)}_{L_{k+1}}(\ux)$. Indeed, by
Corollary 4.1, had box $\BLam^{(N)}_{L_{k+1}}(\uy)$ been both $E$-CNR and $(2m_0)$-NT,
it would have been $(E,m_{k+1})$-NS, which is not allowed within the event $\cB$.
Thus, the box $\BLam^{(N)}_{L_{k+1}}(\ux)$ must be $E$-CNR, but $(E,m_{k+1})$-S:
$$
\cB\cap \cT^{\rm c} \cap \cR^{\rm c} \subset \{\exists \, E\in I:\;\BLam^{(N)}_{L_{k+1}}(\ux)
\text{ is } (E,m_{k+1}){\text{-S}} \text{ and } E{\text{-CNR}} \}.
$$
However, applying Lemma \ref{J_NS}, we see that
$$
\begin{array}{r}
\{\exists \, E\in I:\,
\BLam^{(N)}_{L_{k+1}}(\ux) \text{ is }(E,m_{k+1}){\text{-S}} \text{ and }
E{\text{-CNR}}\}\qquad\qquad\\
\subset \{\exists \, E\in I:\, K(\BLam^{(N)}_{L_{k+1}}(\ux);E) \geq J+1\}.
\end{array}
$$
Therefore, with the same values of parameters
as in Corollary \ref{CorHowManySub},
$$
\begin{array}{cl}
\pr{\cB\cap \cT^{\rm c}\cap \cR^{\rm c}}
&\leq \pr {\exists \, E\in I:\, K(\BLam^{(N)}_{L_{k+1}}(\ux)
;E) \geq 2\ell+2 }\\
\;&\leq 2L_{k+1}^{-1}\,L_{k+1}^{-2p(N)}.\end{array}\eqno (5.4)$$
Finally, we get, with $q'(N):=q(N)/\alpha$,
$$\begin{array}{cl}
\pr{\cB} &\leq \pr{\cB\cap \cT} + \pr{\cR}  + \pr{\cB\cap \cT^{\rm c}
\cap \cR^{\rm c}}\\
\;&\leq \half L_{k+1}^{-2p(N)} + L_{k+1}^{-q'(N)+4} + 2L_{k+1}^{-1} \, L_{k+1}^{-2p(N)}
\leq L_{k+1}^{-2p(N)},\end{array}
\eqno (5.5)
$$
for sufficiently large $L_0$, if we can guarantee, by taking $|g|$ large enough, that
$q'(N) > 2p(N) + 5$. This completes the proof of Theorem \ref{ThmINISing}.
$\qquad\qed$

\psn
{\bf Remark.} The proof of Theorem \ref{ThmINISing} practically repeats that of Theorem 5.1 from
\cite{CS09}; the only difference is in specification of constants in the exponents.
\psn
Therefore, Theorem \ref{MSAInd} is also proven.

\section{Appendix A}

\pn
\textit{ Proof of Lemma \ref{CondGeomSep}}.  Consider two $N$-particle configurations $\ux$ and $\uy$ and introduce the following notion: we shall say that
 the set of positions $\{x_j, j\in \cJ\}$, $\cJ\subseteq \{1,\ldots ,N\}$, form an $R$-connected cluster (or simply an $R$-cluster)
iff the set
$$
\bigcup_{j\in\cJ} \Lam_R(y_j) \subset \Z^d
\eqno (6.1)
$$
is connected. Otherwise, this set of particles is called $R$-disconnected, in which case it can be decomposed into two or more
$R$-clusters. Now, we proceed as follows.
\psn
{\bf (1)} Decompose the configuration $\uy$ into $L$-clusters (of diameter $\leq 2NL$).
\pmn
{\bf (2)} To each position $y_j$ there corresponds precisely one cluster, denoted by $\Gamma (j)$.
Let $\Y =\{\Gamma (j): \; j\in\cJ\}$ stand for the collection of clusters, with $\card \Y \leq N$.
\psn
{\bf (3)} Consider any of the clusters $\Gamma (j) \in \Y$. By definition, $\Gamma (j)$ is disjoint from all other clusters:
$$
\Gamma (j) \cap \Gamma (i) =
\begin{cases}
   \Gamma (j), & \text{ if } \Gamma (i) = \Gamma (j),\\
   \emptyset, & \text{ otherwise}.
\end{cases}
\eqno (6.2)
$$
Therefore, for any two distinct clusters $\Gamma', \Gamma''\in \Y$, the respective sigma-algebras
${\mathfrak B}(\Gamma '), {\mathfrak B}(\Gamma '')$ are independent.
\psn
{\bf (4)} Suppose that $\exists\, j\in\{1,\ldots ,N\}:\, \Gamma (j) \cap \Pi \BLam^{(N)}_L(\ux) = \emptyset$.
Set
$$
\bar{\mathfrak B}_j(\uy) := {\mathfrak B}\left( \cup_{\Gamma (i) \neq \Gamma (j)} \Gamma (i) \right).
$$
Then the sigma-algebra ${\mathfrak B}(\Gamma (j))$ is independent of ${\mathfrak B}(\BLam^{(N)}_L(\ux))$
and of $\bar {\mathfrak B}_j(\uy)$:
$$
{\mathfrak B}(\Gamma (j)) \coprod {\mathfrak B}(\BLam^{(N)}_L(\ux)),
\;{\mathfrak B}(\Gamma (j)) \coprod \bar {\mathfrak B}_j(\uy).
\eqno(6.3)
$$
In other words, the box $\BLam^{(N)}_L(\uy)$ is separable from $\BLam^{(N)}_L(\ux)$.
\psn
{\bf (5)} Suppose (4) is wrong, and let's deduce from the negation of (4)
a necessary condition on possible locations
of the configuration $\uy$, so as to show that the number of possible choices is finite.
Indeed our hypothesis reads as follows:
$$
\forall\, j\in \{1,\ldots ,N\}\;\; Y(j) \cap \Pi \BLam^{(N)}_L(\ux) \neq \emptyset.
\eqno(6.4)
$$
Therefore,
$$
\begin{array}{l}
\forall\, j\in \{1,\ldots ,N\}\;\; \exists\, i:\; \|y_j - x_i\| \leq 4NL+L = (4N+1)L \leq 5NL \\
\Rightarrow \forall\, j\in \{1,\ldots ,N\}\;\; y_j \in \Pi \BLam^{(N)}_{AL}(\ux), \; A  = A(N) = 5N.
\end{array}
$$
We see that if a configuration $\uy$ is not separable from a  $\ux$, then
every position $y_j$ must belong to one of the boxes $\Pi_i \BLam^{(N)}_{AL}(\ux)
= \Lam_{AL}(x_i)\subset\Z^d$.
The total number of these boxes is bounded by $N$.
There are at most $N^N/N!$ choices for the $N$ positions
$y_1, \ldots, y_N$. For any given choice among $J(N) \le N^N/N!$ possibilities,
the point $\uy = (y_1, \ldots, y_N)$ must belong to the Cartesian
product of $N$ boxes of size $AL$, i.e. to an $Nd$-dimensional box
of size $AL$.  The assertion of Lemma \ref{CondGeomSep} now follows. \qed

\section{Appendix B. Finite-volume localisation bounds}

 Here we give the proof of Lemma \ref{GF_rep}. Recall, we consider  operator $\uH^{(N)}_{\BLam_{L_k}(\uu)}$ in a box $\BLam^{(N)}_{L_k}(\uu)$. Let $\BPsi_j$,
$j=1, \ldots, |\BLam^{(N)}_{L_k}|$, be its normalised EFs and $E_j$ the respective EVs. Fix $j$
and consider the GFs $\uG^{(N)}(\uv,\uy;E_j)$, $\uv,\uy\in \BLam^{(N)}$.


\psn
\textit{Proof of Lemma \ref{GF_rep}}. Recall that the CNR property implies NR. Observe that
$E-\lam_a - \mu_b = \left(E-\lam_a \right) - \mu_b$.  Further, by the hypothesis of the lemma, $\BLam^{(N)}_{L_k}(\uu)$ is $E$-CNR. Therefore, for all $\lam_a$, the $n''$-particle box $\BLam^{(n'')}_{L_k}(\uu'')$ is
$(E-\lam_a)$-NR. By the assumption of $m$-NT, $\forall\, E\in I$ box
$\BLam^{(n'')}_{L_k}(\uu'')$ must not contain two disjoint $(E-\lam_a, m)$-S sub-boxes of size $L_{k-1}$.
Therefore, the MSA procedure proves that $\BLam^{(n'')}_{L_k}(\uu'')$ is $(E-\lam_a)$-NS, yielding the required upper bound.

Let us now prove the second assertion of the Lemma.  If $\uv=(\uv',\uv'')\in\pt \BLam^{(N)}_{L_k}(\uu)$, then either $\|\uu' - \uv'\|=L_k$, or
$\|\uu'' - \uv'' \|=L_k$. In the former case we can write
$$
\begin{array}{cl}
 \uG^{(N)}(\uu, \uv;E) &
= \sum_{a} \ffi_a(\uu') \ffi_a(\uv') \, \sum_{b} \, \frac{ \psi_b(\uu'') \psi_b(\uv'') }{ (E - \lam_a) - \mu_b } \\ \\
\;&= \sum_{a} \ffi_a(\uu') \ffi_a(\uv') \,  \,  \uG^{(n'')}_{\BLam^{(n'')}_{L_k}(\uu'')}(\uu'', \uv''; E - \lam_a).\end{array}
\eqno(7.1)
$$
Since $\|\ffi_a \|=1$, we see that
$$
|  \uG^{(N)}(\uu, \uv;E) |
\leq \left| \Lam^{(n')}_{L_k}(\uu') \right| \,
\max_{\lam_a}\, | \uG^{(n'')}_{\Lam^{(n'')}_{L_k}(\uu'')}(\uu'', \uv''; E - \lam_a)|.
\eqno(7.2)
$$
In the case where $\|\uu'' - \uv'' \|=L$, we can use the representation
$$
 \uG^{(N)}(\uu, \uv;E)
= \sum_{b} \psi_b(\uu'') \psi_b(\uv'') \,  \,  \uG^{(n')}_{\Lam^{(n')}_{L_k}(\uu')}(\uu', \uv'; E - \mu_b).
\eqno(7.3)
$$
\qed

Now, as was said before, Lemma \ref{lem_one} follows from Lemma \ref{GF_rep} combined with the bounds
\dsk{n'}, \dsk{n''}, for $1 \le n', n''<N$.

\pmn{\bf  Acknowledgments.}
VC thanks The Isaac Newton Institute (INI) and Department of Pure Mathematics and Mathematical Statistics,
University of Cambridge, for hospitality during visits in 2003, 2004,
2007 and 2008. YS thanks the D\'{e}partement de Math\'{e}matiques,
Universit\'{e} de Reims for hospitality during visits in
2003 and 2006--2008, in particular, for a Visiting Professorship
in the Spring of 2003. YS thanks IHES, Bures-sur-Yvette, and
STP, Dublin Institute for Advanced Studies, for hospitality
during visits in 2003--2007. YS thanks the Departments of Mathematics
of Penn State University  and of UC Davis, for hospitality
during Visiting Professorships in the Spring of 2004, Fall of
2005 and Winter of 2008.  YS thanks the Department of Physics,
Princeton University and the Department of Mathematics of UC Irvine,
for hospitality during visits in the Spring of 2008. YS
acknowledges the support provided by the ESF Research Programme
RDSES towards research trips in 2003--2006.

We are grateful to the INI, where our multi-particle project was originated in 2003,
during the programme \textit{Interaction and Growth in Complex Systems}. Special thanks go to the organisers of the 2008 INI programme \textit{Mathematics and Physics of Anderson Localization: 50 Years After}; for our project the mark
  became  \textit{First Five Years After}.

\bibliographystyle{amsalpha}

\end{document}